\documentclass[12pt,amsfonts]{article}
\usepackage{amsfonts}

\def\half{\textstyle{\frac{1}{2}}}
\def\quarter{\textstyle{\frac{1}{4}}}

\def\H{{\cal H}}

\def\threebytwo{\textstyle{\frac{3}{2}}}
\def\eigth{\textstyle{\frac{1}{8}}}

\def\p{\phi}
\def\H{{\cal H}}

\def\l{\lambda}

\def\S{\Sigma'}
\def\t{\textstyle}

\def\F{{\cal F}}

\def\ra{\rightarrow}
\def\tint{{\textstyle\int}}

\def\hp{{\hat\pi}}
\def\hph{{\hat\phi}}
\def\s{\hskip.08em}

\def\d{\partial}

\def\a{\alpha}
\def\b{\begin{eqnarray*}}  
\def\e{\end{eqnarray*}}    
\def\bn{\begin{eqnarray}}  

\def\<{\langle}
\def\>{\rangle}

\def\no{\nonumber}

\def\k{\kappa}
\def\{{\lbrace}
\def\}{\rbrace}
\title{Rethinking Renormalization}
\author{John R. Klauder\footnote{klauder@phys.ufl.edu}
\\Department of Physics and \\Department of Mathematics\\
University of Florida\\
Gainesville, FL 32611-8440}
\date{ }
\begin{document}
\maketitle
\begin{abstract}
As applied to quantum theories, the program of renormalization is
successful for `renormalizable models' but fails for
`nonrenormalizable models'. After some conceptual discussion and
analysis, an enhanced program of renormalization is proposed that is
designed to bring the `nonrenormalizable models' under
control as well. The new principles are developed by studying several,
carefully chosen, soluble examples, and include a recognition
of a `hard-core' behavior of the interaction and, in special cases,
an extremely elementary procedure to remove the source of all
divergences. Our discussion provides the background for
a recent proposal for a nontrivial quantization of nonrenormalizable
scalar quantum field models, which is briefly summarized as well.\vskip.3cm
{\bf Dedication}: It is a pleasure to dedicate this article to the memory of
Prof.~Alladi Ramakrishnan who, besides his own important contributions to science,
played a crucial role in the development of modern scientific research and education
in his native India. Besides a number of recent informative discussions during his yearly
visits to the University of Florida, the  present author had the pleasure much
earlier of
hosting Prof.~Alladi during his visit and lecture at Bell Telephone Laboratories.
\end{abstract}

\section*{Introduction}
Renormalization has been a very successful paradigm for dealing with
an important class of quantum theories. Its basic principles are
easily stated: The parameters of a classical theory are different
from those of a quantum theory because of additional self interaction
that arises in a quantum theory.
In practical terms, the interacting system is commonly treated as a
perturbation of a free system, and the power series in the nonlinear
coupling often displays divergent terms that need to be canceled and
counterterms of a suitable nature are introduced to do just this. If
a finite number of distinct counterterms can be found so that every
term in the power series expansion is rendered finite, then the
theory is called renormalizable, and many such theories have had
highly successful applications and in several cases have led to
astonishingly accurate predictions when compared to experimental
measurements. This aspect of the program of renormalization is
considered to be a resounding success and deservedly
so. It is natural of course that a successful program such as
renormalization has also been proposed to study a wider class of
theories than its proponents originally intended, and this is indeed
the case. A certain family of field theories fall into the class of
being ``nonrenormalizable", an attribute that asserts that the
procedures usually ascribed to the program of renormalization are
unsuccessful in dealing with certain model problems. If such
examples were confined to esoteric models with no potential
application to the real world, it would be permissible to ignore those
models that are classified as nonrenormalizable. But that
is not the case. The most famous example corresponds to the Einstein
gravitational field for which the general consensus is that quantum
gravity is perturbatively nonrenormalizable. Since the standard
procedures of renormalization have failed for such an important
case, there have been proposed elaborate alternative  theories that
entail additional fields or degrees of freedom that are designed to
produce a theory that is term-by-term finite within a perturbation
analysis. Superstring theory is one such program, and $N=8$ supergravity
is another. In so doing, these alternative theories have
introduced additional fields, which, thanks to the differing properties
of fermions and bosons can lead to cancellations among the old,
divergent contributions of the original theory and well chosen, new,
divergent contributions from the carefully selected additional fields. This general
approach is sufficiently broad that it would seem to cover all
possible situations regarding how interactions and auxiliary counterterms can
appear and interact with each other. However, there is one important
class of models that is in practice not covered by the preceding
characterization. Admittedly, it is not obvious where one should
look if such an overlooked class of examples is to be found. A
clue to the overlooked class emerges if we recall that the
traditional procedures of regularization and renormalization entail
the implicit assumption that if the perturbative interaction is
reduced in strength, say by the usual device of reducing the value of
the associated coupling constant,  then, in the limit that the
coupling constant vanishes and the effect of the interaction is
formally eliminated, the resulting theory in the limit of a
vanishing coupling constant is identical to the free theory with
which one started. Stated otherwise, and perhaps more directly,
this is the implicit assumption that the set of interacting
theories defined as the set that is produced for all nonzero
(typically positive) values of the coupling constant is such that as
the coupling constant goes to zero, the limit of that set of
interacting theories is the free theory itself, i.e., the
interacting theories are {\it continuously connected to the free
theory}. This highly natural, implicit assumption covers a lot of the
important cases but it certainly does not cover all possibilities
some of which may have some ultimate physical relevance. It is an
important feature of this paper that we focus on these outlier model theories,
which are typically nonrenormalizable models.
\subsection*{Overview of the Present Paper}
The features ascribed to the renormalization program are not limited
to quantum field theory but also arise in quantum mechanical analogues.
As such, one can gain real insight into the distinction among super
renormalizable, strictly renormalizable, and nonrenormalizable models.
A common feature of the latter theories is the occurrence of a {\it hard-core
potential}. From a (Euclidean) functional integral viewpoint, the nonlinear
interaction acts partially as a hard core projecting out certain paths that
would otherwise appear in the free theory. This fact -- which we believe
is a defining characteristic for a large class of nonrenormalizable interactions --
means that an interacting theory is {\it not} continuously connected to
the free theory as the coupling constant is reduced to zero. This property of the
quantum theory is also seen in the classical theory itself by the fact that, generally
speaking, the set of solutions of the interacting classical theory does not reduce
to the set of solutions that characterizes the free solutions. This aspect will be
illustrated for particle systems as well as field systems.

The full dynamics of a classical system involves the action functional and its
stationary variation to derive the equations of motion. In a (Euclidean) functional
integral formalism, the classical action again plays an important role in the
quantum dynamics. Regularization is essential in order to give a functional
integral meaning, and it is customary to use a lattice approximation for the time
for particle mechanics or for spacetime for field models. The lattice
action induces a lattice Hamiltonian operator and in turn a lattice ground state
for that Hamiltonian. It is natural that a model can be characterized by either the
action, the Hamiltonian, or the ground state. It is important to remark that we
focus heavily on the ground state in our analysis.
When we take up the discussion of field problems, we will present an argument
that shows an important role that the ground state plays.

However, before dealing with fields, we wish to illustrate how the issue of
renormalization arises in elementary one dimensional examples.

\section*{One Dimensional Example}
 Consider a
classical system for a single, phase space, degree of freedom
$(p,q)$ with a classical Hamiltonian given by
    $$ H_\lambda(p,q)=  \half( p^2 +q^2) +\lambda |q|^{-\alpha}\;. $$
For any $\alpha>0$, it follows, just from energy considerations,
that the motion of the particle can never be such as to reach the
origin $q=0$ let alone pass through the value $q=0$. This situation
holds for all values of the coupling constant $\lambda>0$,  and as a
consequence, as $\lambda\ra0$, the set of classical solutions of the interacting
theory do {\it not} correspond to the set of classical solutions of the free theory,
namely, that of the free harmonic oscillator given by $q(t)=A
\cos(t-a)$. Specifically, for any choice of the amplitude $A$ and
the phase $a$ there will be for
{\it every} solution of the free theory a time $t$ for which the
solution vanishes and even crosses the line $q=0$. In contrast, the
solutions of the interacting theory for which $\lambda>0 $, all pass
by continuity to solutions not of the free theory but to those which
are rectified in the sense that they are of the form  $q(t)=
\pm|A\cos(t-a)|$ and are all strictly different over time from the
usual free theories. We give the name {\it pseudofree} to the name
of the theory, different from the free theory, to which the
interacting theory is continuously connected as the nonlinear
coupling constant goes to zero. Clearly, if one reintroduces the
interaction starting from the pseudofree theory, the form of the new
solutions is indeed continuously connected to that of the pseudofree
theory.

The easiest way to characterize the pseudofree  quantum theory is by
its Hamiltonian which is the same as that of the free harmonic
oscillator augmented by Dirichlet boundary conditions at $x=0$.  If
one were contemplating a perturbation series representation of the
interacting solution, that power series should not be about the free
theory (to which the interacting solutions are not continuously
connected!) but rather about the pseudofree theory.

Regarding the quantization of such a model, there are some surprises
that can arise. For example, when $0<\alpha<1$, it follows that the
interacting quantum solution is in fact continuously connected to
the free quantum theory unlike the situation for the classical case.
For $\alpha>2$, on the other hand, there is no modification of the
theory that can be made to prevent the theory from passing to a
pseudofree theory as the parameter $\lambda\ra0$. In other words,
for $\alpha>2$, the interacting quantum theory passes to a
pseudofree theory with a set of eigenfunctions and eigenvalues that
are generally different from those that characterize the free theory. What
happens in the interval $1\le\alpha\le2$ is quite interesting and to
some extent open to different conclusions. With an eye toward
maintaining a continuous connection of the interacting theories to
the free theory, it is possible to choose a regularized form for the
interaction, namely, a set of potentials of the form
$V_\epsilon(q,\l)$ that have the property that as $\epsilon\ra0$,
the regularized potentials
    $$ V_\epsilon(q,\l)\ra \l\s |q|^{-\alpha}\;, \hskip2cm q\neq0\;.$$
These regularized forms of the potential are rather strictly
constrained and they involve polynomial contributions in the
coupling constant $\lambda$. It is not difficult to determine the
general form of the regularized potential simply on the basis of
dimensional arguments. In particular, the dimensions of the Hamiltonian
are those of the first term $p^2$, and taking Planck's constant
$\hbar=1$ for the present time, the dimensions are that of $L^{-2}$ where
$L$ denotes the dimension of length. With the regularization
parameter $\epsilon>0$ entering initially in the interaction as
    $$ \l\s|q|^{-\alpha}\ra \l\s(|q|+\epsilon)^{-\alpha} \;,$$
it follows that the dimension of $\epsilon$, like $q$, is $L$. In order that
the interaction terms have the right dimensions, i.e., $L^{-2}$, it
follows that the dimension of $\lambda$ is that of $ L^{\alpha-2}$.
For regularization terms we restrict ourselves to terms of the form
     $$ k_j\s\lambda^j\,\epsilon^{-p_j}\,\delta(q)\;,$$
     where $\delta(q)$ is a Dirac delta function.
With $k_j$ chosen as an unknown dimensionless factor, and since
$\delta(q)$ has dimensions $L^{-1}$, it follows that the power
$p_j=1-(2-\alpha)j$ in order to ensure that the regularization terms
above each have the desired dimension of the Hamiltonian, namely
$L^{-2}$.  Hence the regularized form of the potential is given by
   $$V_\epsilon(q,\l)=\l\s(|q|+\epsilon)^{-\alpha}-\sum_{j=1}^J\,k_j\s\,\lambda^{j}\,
   \epsilon^{(2-\alpha)j-1}\,\delta(q)\;.$$
The factor $J$ denotes the upper limit of the sum which occurs
whenever $(2-\alpha)^{-1}$ is nonintegral and $(2-\alpha)J<1<(2-\alpha)(J+1)$ for then all further regularization
terms vanish as $\epsilon\ra0$. In this case  further analysis shows that the factors $k_j$ are
given by $k_1=2/(\a-1)$ and then
    $$  k_{j}=-\frac{1}{[1-j(2-\a)]}\sum_{q=1}^{j-1}\s k_{j-q}\s k_q \;; $$
if instead, $(2-\alpha)^{-1}=J$ is an integer, then
the last factor $k_J$ involves a natural logarithm; see \cite{book}.
For  $\alpha=2$, $J=\infty$, and all $p_j=1$. For all $\alpha\le2$
such a series provides a regularized potential for which the
interacting theory is continuously connected to the free theory as
$\lambda\ra0$. It is noteworthy that when $\alpha<2$ a finite series
of counterterms, each with a  diminishing divergence (i.e.,
$p_{j+1}<p_j$), provides the proper regularized potential, a
property similar to that encountered when dealing with super
renormalizable quantum field theories. When $\alpha=2$ an infinite
series of counterterms, all of equal divergence (i.e.,
$p_{j+1}=p_j$), leads to a suitable regularized potential, a property
similar to that of so-called strictly renormalizable quantum field
theories. For $\alpha>2$, on the other hand, there is no regularized
potential that leads to an interacting theory that is continuously
connected to the free theory.  Of course, the proposed regularization
terms based simply on dimensionality do not know this fact, and it
may be said that they do their best to signal their inability to
provide a solution to the problem by the fact that when  $\alpha>2$,
the term-by-term divergence actually {\it increases} (i.e., $p_{j+1}>p_j$),  and
moreover,  $p_j\ra \infty$ as $j\ra\infty$, a property which is
reminiscent of the behavior of nonrenormalizable quantum field
theories.
\subsubsection*{A brief summary}
We have discussed this simple quantum
mechanical model in some detail in order to show what kind of
singular behavior is possible even in quantum mechanics.  In
particular, we observe that for $\alpha<1$, there is no anomalous
behavior in the quantum theory although there is anomalous classical
behavior. For $1\le\alpha\le2$, it can be arranged that there is no
anomalous  quantum behavior although there always will be anomalous
classical behavior. The price to pay for this good quantum behavior
is the introduction of regularized quantum terms that entail a power
series in the coupling constant $\lambda$. For $\alpha>2$, on the
other hand, there is no escaping the anomalous quantum behavior no
matter how one tries to regularize the quantum theory.
\subsubsection*{Field theory analog -- a brief detour}
 We claim there is an analog
with the above story for quantum mechanics that plays out in quantum
field theory as well. For sufficiently weak perturbations, the
interaction can be renormalized so that the resultant interacting
theory is continuously connected to the free theory as the coupling
constant is reduced to zero; this is the situation that applies to
super renormalizable and possibly to strictly renormalizable theories.
For sufficiently strong perturbations, the interaction cannot be
renormalized so that the interacting theory is continuously connected
to the free theory. Instead, for such strong perturbations, the
interacting theories are connected to an appropriate pseudofree
theory. Later, we will bolster the argument that this is the
situation which should apply to nonrenormalizable theories. To make
this leap of faith from a singular family of classical problems and
their associated  quantum problems to a wide class of quantum field
theories, it will be helpful to develop a primary {\it principle} that
captures the essence of the singular nature of the interaction that
leads to either a continuous connection with the original free
theory or instead leads to a continuous connection with a pseudofree
theory.
\subsection*{Path Integral Formulation}
The principle we adopt to describe the appearance of pseudofree
theories is that of a {\bf hard-core
interaction}.  The concept behind this principle is most simply
appreciated in a functional integral representation of the
associated quantum system. This analysis works for either a real
time or an imaginary time functional integral, and for its better
mathematical structure, we shall choose the latter form.  For the
quantum mechanical problem that we have so far been discussing, the
associated imaginary time (Euclidean) functional integral is given
by
   $$ {\cal N}\int e^{\t-\tint\{{\half }\s[{\dot x}^2+x^2]+\lambda\s V(x)\} \,dt}\;{\cal D}x\;. $$
Although the Brownian-like paths $x(t)$ that enter this functional
integral have a nowhere defined (i.e., divergent) derivative -- a
feature that is surely unlike the classical theory -- it is
noteworthy that the distinction between the behavior for $\alpha<2$
and $\alpha>2$ can nevertheless be won by simple classical
arguments. For classical paths consider the following simple
inequality
    $$|x(t_2)-x(t_1)|=|\int_{t_1}^{t_2}\s {\dot x}(t)\,dt|\le |t_2-t_1|^{1/2}\,
    \bigg[{\int_{t_1}^{t_2} {\dot x}^2(t)\,dt}\bigg]^{1/2}\;. $$
Assuming a finite value for the kinetic energy, it follows, for some
$K<\infty$, that
   $$ |x(t_2)-x{t_1}|^{-\alpha}\ge K \,|t_2-t_1|^{-\alpha/2}\;. $$
Setting $x(t_2)=0$, the location of the singularity, we see that
    $$ \int|x(t)|^{-\alpha}\,dt\ge K\,\int |t|^{-\alpha/2}\,dt\;. $$
This inequality implies that for $\alpha >2$ the integral over the
interaction term diverges, while for $\alpha<2$ that is not
necessarily the case.  When the integral over the interaction
diverges, the contribution of that path is projected out (by the
factor $e^{-\infty}$) for any positive value of the coupling
constant. And as the coupling constant is reduced to zero, the
contribution of that path is never restored leading to the exclusion
of that path in the definition of the pseudofree theory. For the
quantum mechanical  problem previously discussed, this means that
whenever $\alpha>2$, the contribution of all paths that reach or
cross the axis $x=0$ are projected out of the functional integral;
that is the meaning of the statement that the interaction acts in
part like a hard core. Our simple argument involving the inequality
derived from classical paths does not have anything to say about
what happens for $\alpha<2$, but that does not diminish its
importance for the region $\alpha>2$.

Before proceeding, let us restate some important issues that arose
in our analysis of the one dimensional quantum problem as discussed
above. The model we studied had a clearly defined free theory (with
$\lambda\equiv0$) which is just the usual harmonic oscillator. The
free propagator (in imaginary time for convenience) is readily given
by the sum
  $$\<x'',T|x',0\>=\sum_{n=0}^\infty \,h_n(x'')\,e^{-(n+1/2)T}\,h_n(x')\;, $$
where the   set of functions $\{h_n(x)\}_{n=0}^\infty$ are the
Hermite functions defined by the generating function
    $$\exp(-s^2+2sx-{\half}\s x^2)=\pi^{1/4}\,\sum_{n=0}^\infty\,(n!)^{-1/2}\,(s\s\sqrt{2})^n\,h_n(x)\;.$$
In the present case the pseudofree theory (denoted by a prime $'$) has a propagator defined
by the expression
  $$\<x'',T|x',0\>'=\theta(x''\s x')\sum_{n=0}^\infty \,h_n(x'')\,e^{-(n+1/2)T}\,[\s h_n(x')-h_n(-x')\s]\;, $$
where the function $\theta(u)=1$ if $u>0$ and $\theta(u)=0$ if
$u<0$. It is the latter expression that incorporates the hard core,
projecting out all those paths in the free harmonic oscillator
propagator that reach or cross the value $x=0$. Note well: It is the
pseudofree theory to which the interacting theories are continuously
connected as the coupling constant is reduced to zero. It is the
pseudofree theory around which a meaningful perturbation theory for
the singular perturbation can be constructed. From the point of view
of a Euclidean functional integral, if one attempted to expand a
partially hard core interaction about the free theory, this would
lead to a series composed of ever more divergent expressions.
Regularization of that series would serve to render those terms
finite but it would also falsely imply that the interacting theory
was continuously connected to the free theory because the
regularized power series would reduce to the free theory when the coupling
constant is reduced to zero.  This property of the regularized
perturbation series is entirely erroneous and misleading.

Moreover, the seed of the discontinuous nature of the perturbation
about the free theory is already evident in the classical theory
itself. This situation holds because the classical solutions of the
interacting theory already do not reduce to the solutions of the
classical free theory as $\lambda\ra0$. Instead they pass to the
classical solutions of the pseudofree theory as noted above. This
result has the important consequence that an indelible imprint of
the fact that one could be dealing with a discontinuous perturbation
(of the free theory) can be determined from an analysis of the
classical interacting theory itself! The nature of such an analysis
is not too difficult; it rests on the determination that the
set of solutions of the interacting theory for arbitrarily small
coupling constant is not equivalent to the set of solutions of the
free theory itself.

The criterion that a classical pseudofree theory be different from the
classical free theory is necessary for a quantum pseudofree theory to be
different from a quantum free theory. However, the one dimensional example
with $0<\a<1$ demonstrates that such a criterion is not sufficient to ensure that
the quantum  theory  also involves a pseudofree theory different from the
free theory.

\subsubsection*{Shifting the singularity from $x=0$ to $x=c$}
Suppose, instead of the singularity being at $x=0$, we moved it to
the point $x=c$, where without loss of generality we can assume that
$c>0$. This means that our basic potential is $\l\s|x-c|^{-\alpha}$. We
now briefly summarize the main changes that occur. First, the
classical story.  In this case, the free solution given by
$q(t)=A\cos(t-a)$ may remain unchanged if the overall classical
energy is sufficiently small, which occurs when $|A|\le c$. When
$|A|>c$, two solutions are possible, one of the form
$q(t)=\max[A\cos(t-a),c]$ with the phase $a$ adjusted so that
the classical path continues to obey the equation of motion. The second
path is given by $q(t)=\min[A\cos(t-a),c]$ with the phase again
adjusted so that the classical path solves the equation of motion.
The quantum theory for this case is such that the pseudofree
theory is defined by the harmonic oscillator Hamiltonian augmented
by Dirichlet boundary conditions at $x=c$. As a consequence, the
eigenfunctions and eigenvalues of the free harmonic oscillator are
almost never relevant in the construction of the pseudofree
Hamiltonian.  The same conclusions would be drawn from an analysis
of the Euclidean functional integral formulation of the quantum
theory.  For $\alpha\le2$, a regularized potential qualitatively
similar to that discussed before, should be suitable to define an
interaction that is continuously connected to the free theory. For
$\alpha>2$, however, no regularized form of the potential leads to
interacting theories that are continuously connected to the free
theory as the coupling constant passes to zero. Any perturbation
analysis of the interacting theory when $\alpha>2$ must take place
about the pseudofree theory. It is noteworthy in this example that
as $c\ra\infty$ the classical solutions all tend to those of the
free theory. It is also true that as $c\ra\infty$, the
pseudofree quantum theory passes to the free quantum theory.
\subsubsection*{A remark on higher dimensional examples}
 Although these facts have
been illustrated for a comparatively simple one-dimensional
classical/quantum model, it is not difficult to imagine analogous
situations in higher dimensional mechanical systems that lead to a
corresponding behavior. For example, a two-dimensional configuration
space may have a singular potential of the form
$\l\s(x^2+y^2)^{-\alpha}$. However, this example does {\it not} lead to
a discontinuous perturbation since, although there are Brownian
motion paths that pass through the singular point $x=y=0$ and which
therefore need to be discarded, the set of such paths is only of
measure zero. To achieve a discontinuous perturbation, one would
need a singularity of co-dimension one such as offered by the
potential $\l\s|(x^2+y^2)-1|^{-\alpha}$, for example. There is a rich
set of examples of this sort, but we shall not dwell on them for we
are after still bigger game, namely, those that arise for an
infinite number of variables!

\section*{Classical and Quantum Field Theory}
Until now, we have seen  simple models for which the interacting theory is not continuously connected
to the free theory as the coupling constant is reduced to zero. In the classical
regime, such a situation can be seen by comparing the set of solutions allowed by the
free classical theory with the set of solutions allowed by the pseudofree classical theory.
In those cases where the set of solutions of the pseudofree classical theory is a proper
subset of the set of solutions of the free classical theory, we have a genuine situation where the
interacting theory has left an indelible imprint on the classical theory as the coupling constant
is reduced to zero. When it comes to an analysis of the associated quantum theories, however,
the classical results offer only a partial guide. In certain cases, the interacting quantum theory
is continuously connected to the free theory, and thus there is no distinct pseudofree quantum theory,
even though the classical pseudofree and free theories differ from one another; for example, this is
the case for the one dimensional model when $0<\alpha<1$.  In such a case,
it is natural that a quantum perturbation series about the free theory would be the proper
choice. However, there is still another option, and this is the one to which we wish to draw
attention, namely when the pseudofree quantum theory is distinct from the free quantum theory.
It is for such situations that the interacting quantum theory is not continuously connected to the free
quantum theory
as the coupling constant is reduced toward zero. It is in such cases that a perturbation series of the
interaction taken about the free theory would be wrong while a perturbation series about the
pseudofree theory would be the proper choice; for example, this is the case for the
one dimensional models when $\alpha>2$.

          \subsection*{Focus on the Ground State}
          We aim to carry these concepts from one dimensional systems to field theoretic systems. Functional
integral formulations entail regularization such as that offered by a lattice.

Consider the spacetime lattice formulation of a general problem phrased as a
scalar field theory. Let $\p_k$ denote the field value at the lattice point
$k=(k_0,k_1,k_2,\ldots,k_s)$, where $k_j\in \{0,\pm1,\pm2,\ldots\}\equiv{\mathbb Z}$, $k_0$ refers to the
(future) temporal direction, and the remaining $k_j$, $1\le j\le s$, denote the
$s$ spatial directions; for a quantum mechanical problem, $s=0$. Assume that spacetime
is replaced by a periodic, hypercubic lattice with $L$ points on an edge and $L^s\equiv N'$ lattice points in a spatial slice.

             In this section we first wish to argue that moments of expressions of interest in the
full spacetime distribution can be bounded by suitable averages of related quantities in the
ground state distribution. In particular, let the full spacetime average on a lattice be given by
     $$ \<\s[\Sigma_{k_0}\s F(\p,a)\s a\s]^p\s \>\equiv
     M\int [\Sigma_{k_0} F(\p,a)\s a\s]^p\;e^{\t -I(\p,a,\hbar)}\;\Pi_k\s d\p_k \;, $$
     where $I$ is the lattice action, $\Sigma_{k_0}$ denotes a summation over the temporal direction $k_0$ only, and $F(\p,a)$ is an
     expression that depends only on fields $\p_k$ at a fixed value of $k_0$. For example,
     one may consider $F(\p,a)=\Sigma'_k \p_k^4\s a^s$ or $F(\p,a)=\Sigma'_{k,l}\,\Omega_{k,l}\s\p_k\s\p_l\s a^{2s}$, for some $c$-number kernel $\Omega_{k,l}$, etc., where the primed sum implies summation
     over a spatial slice at fixed $k_0$. It follows that
       $$ \<\s[\Sigma_{k_0}\s F(\p,a)\s a\s]^p\s \>=\Sigma_{k_0,\ldots,k_0}\s a^{p}\,\<\s F(\p_1,a)\cdots F(\p_p,a)\s\>\;, $$
       where each $\p_j$ refers to the fields at Euclidean time ``$k_0=j$''. A straightforward inequality shows that
         $$ |\<\s F(\p_1,a)\cdots F(\p_p,a)\s\>|\le |\<\s F(\p_1,a)^p\s\>\cdots \<\s F(\p_p,a)^p\s\>|^{1/p}\;.$$
         Finally, for sufficiently large $N'(ba^s)$, we note that
           $$ \<\s F(\p,a)^p\s\>=\int \s F(\p,a)^p\,\Psi(\p)^2\,\Pi'_k\s d\p_k\;, $$
           namely, an average in the  ground state distribution. The argument behind the last equation is
           as follows. Quite generally,
           $$\<\s F(\p,a)^p\s\>=M{\t\sum}_{l}\int \<\p|\s l\>\s e^{-E_{l}\s T}\s\<l|\s\p\>\s F(\p,a)^p\,\,\Pi'_k\s d\p_k\;, $$
           where we have used the resolution of unity $1=\tint\s|\s\p\>\<\s\p|\,\Pi'_k\s d\p_k$ for states
           for which ${\hat\p}(x)\s|\s\p\>=\p(x)\s|\s\p\>$, as well as the eigenvectors $|\s l\>$ and eigenvalues
           $E_{l}$ for which $\H\s|\s l\>=E_{l}\s|\s l\>$.
           For asymptotically large $T$, it follows that only the (unique) ground state contributes, and the
           former expression becomes
              $$ \<\s F(\p,a)^p\s\>=\int \s F(\p,a)^p\s |\<\p|\s 0\>|^2\, \Pi'_k\s d\p_k\;, $$
              now with $M=1$, which is just the expression given above.

           In summary, for a finite, hypercubic lattice with periodic boundary conditions, we have derived an important result: {\bf If the sharp time average of $[\s F(\p,a)\s]^p$ is finite, then it follows that the spacetime
           average of $[\s\Sigma_{k_0}\s F(\p,a)\s a\s]^p$ is also finite}.

\section*{Ultralocal Scalar Quantum Fields}
As we have done before, we want to illustrate the existence of a pseudofree quantum field theory
distinct from any free quantum field theory by means of a straightforward and soluble example. The example
we have in mind is the so-called {\it ultralocal scalar quantum field theory}. This model has been
rigorously solved previously, and its most complete story can be found in Chap.~10
of \cite{book}. We start with a brief summary of this model based on that rigorous, nonperturbative
analysis. Later we show how a simple and natural argument arrives at a
completely satisfactory solution as well. The advantage of having this simple, alternative argument is
that it can be generalized to realistic, relativistically covariant model quantum field theories.

The classical Hamiltonian for a scalar ultralocal field theory with a quartic nonlinear
interaction is given by
  $$ H=\tint \{\s\half[\pi(t,x)^2+m_0^2\s\phi(t,x)^2]+g_0\phi(t,x)^4\s\}\,d^s\!x\;. $$
  Here, $s$ is the number of spatial dimensions which is one less than the number $n$
  of spacetime dimensions, $s=n-1$. Note well the absence of spatial derivatives in this expression.
  Clearly this is not a relativistic model; rather it is a mathematical model that
will teach us a great deal when it is successfully quantized.

Initially, we note that there are many functions $\p(t,x)$ such that
  $$ \tint[\s{\dot\p}(t,x)^2+m^2_0 \p(t,x)^2\s]\,dt\s d^s\!x<\infty\;,
  \hskip1cm \tint \p(t,x)^4\,dt\s d^s\!x=\infty\;, $$
  a fact which implies  that there is a classical pseudofree theory distinct from the
  classical free theory. This is an important preliminary remark as we try to determine the
  status of the quantum theory.

However, let us first make
a few remarks about the classical properties of such models.
\subsubsection*{Classical features}
The classical equations of motion for this model are given by
  $$ {\ddot\phi}(t,x)+m_0^2\s\phi(t,x)+4\s g_0\s \phi(t,x)^3=0\;. $$
Indeed, the variable $x$ is strictly a spectator variable in this equation,
and we can relegate it to a subsidiary role simply by rewriting the equation of motion as
    $$ {\ddot\phi}_x(t)+m_0^2\s\phi_x(t)+4\s g_0\s \phi_x(t)^3=0\;, $$
which shows the equation of motion is simply that of an independent anharmonic oscillator at each point
of space. Its solution is given by $\phi(t,x)\equiv\phi_x(t)$, where the latter function is
based on the initial data, e.g., $\phi(0,x)\equiv\phi_x(0)$ and ${\dot\phi}(0,x)\equiv{\dot\phi}_x(0)$,
 two functions of $x$ which may be taken to be continuous in $x$, but need not be so.

 Indeed, thanks to the independence of the solution for distinct $x$ values, one may readily discretize this
 model by replacing the spatial continuum by a hypercubic spatial lattice with a lattice spacing $a$
 and $L$ sites on each edge, which leads to a spatial volume give by $V'\equiv (L\s a)^s\equiv N'\s a^s$.
 To begin, we may replace the classical Hamiltonian by a lattice regularized version given by
    $$ H_{reg}={\t\sum'_k}\{\s\half [\pi_k(t)^2+m_0^2\phi_k(t)^2]+g_0\s\phi_k(t)^4\s\}\,a^s\;, $$
    where  $k\in {\mathbb{Z}}^s$;
    this expression is nothing but a Riemann sum approximation to the integral given above, and
    it will converge to the former with $x=\lim\s k\s a$, as the lattice spacing $a$ converges to zero. This regularized Hamiltonian gives rise to the regularized equations of motion
      $$ {\ddot\phi}_k(t)+m_0^2\s\phi_k(t)+4\s g_0\s\phi_k(t)^3=0\;,$$
      and even this set of discrete equations of motion converge to the continuum form
      of the equation of motion as $a\ra0$ and $k \s a\ra x$.
  \subsubsection*{Free ultralocal field theory}
      An important limiting case arises when $g_0=0$ which is the free theory given by
      the free Hamiltonian
         $$ H_0=\half \tint [\pi(t,x)^2+m_0^2\s\phi(t,x)^2\s]\,d^s\!x\;. $$
         The associated free equations of motion are given by
           $$ {\ddot\phi}(t,x)+m_0^2\s\phi(t,x)=0\;, $$
 with a solution given in terms of the initial data $\phi(0,x)\equiv\phi_x(0)$ and ${\dot\phi}
 (0,x)\equiv{\dot\phi}_x(0)$,  by the relation
     $$\phi(t,x)=\phi_x(0)\,\cos(m_0\s t)+m_0^{-1}{\dot\phi}_x(0)\s\sin(m_0\s t)\;, $$
     along with $\pi(t,x)={\dot\phi}(t,x)$, or specifically by
     $$\pi(t,x)=-m_0\s\phi_x(0)\,\sin(m_0\s t)+{\dot\phi}_x(0)\s\cos(m_0\s t)\;. $$
   The lattice regulated free Hamiltonian and the associated free solution is also easily given
  by
     $$H_0=\half\sum'_k[\s \pi_k(t)^2+m_0^2\s\phi_k(t)^2\s]\,a^s\;, $$
     as well as
     \b &&\phi_k(t)=\phi_k(0)\,\cos(m_0\s t)+m_0^{-1}{\dot\phi}_k(0)\s\sin(m_0\s t)\;,\\
      &&\pi_k(t)=-m_0\s\phi_k(0)\s\sin(m_0\s t)+{\dot\phi}_k(0)\s\cos(m_0\s t)\;. \e
    The free model is therefore nothing but an infinite number of identical harmonic oscillators all with
    the same angular frequency $m_0$! Clearly, as $a\ra0$ and $k\s a\ra x$, the regularized solutions
    $\phi_k(t)$ and $\pi_k(t)$ converge to the continuum solutions $\phi(t,x)$ and $\pi(t,x)$.
\subsection*{Quantum Theory -- First Look}
We start the discussion of the quantum theory with the free theory. We promote the classical field
at time $t=0$ (and then suppress the time argument) to an operator field $\phi(x)\ra\hph(x)$ as well
as promote the classical momentum $\pi(x)
\ra\hp(x)$, subject to the canonical commutation relation (in units where $\hbar=1$)
    $$ [\hph(x),\hp(y)]=i\s\delta(x-y)\;.$$
 The free quantum Hamiltonian $\H_0$ is then written as
   $$ \H_0=\half\tint[\s :\hp(x)^2+m_0^2\s\hph(x)^2\s:\s]\,d^s\!x\;, $$
   where, as usual, the notation $:(\s\cdot\s):$ denotes normal ordering (all creation operators to the left
   of all annihilation operators). We denote by $|\s0_0\>$ the nondegenerate ground state of $\H_0$ for which
   $\H_0\s|\s0_0\>=0$ holds, thanks to the normal ordering which removes the (infinite) zero-point energy.

   An important relation that characterizes the ground state eigenstate is the expectation functional
     $$E_0(f)\equiv \<0_0|\s e^{\t i\tint\hph(x)\s f(x)\,d^s\!x}\s|\s0_0\>=e^{\t-(1/4m_0)\tint f(x)^2\,d^s\!x}\;.$$
     Indeed, the structure of this functional as the exponential of a local integral of $f(x)$ is dictated by the fact that the temporal development of the operators at any point $x$ is ultralocal, i.e., the
     temporal development at $x$ is completely independent of
     the time development at a different spatial point $x'$. This behavior carries over to the case of the
     interacting ultralocal model as well, and one expects that whatever the full Hamiltonian operator $\H$ is, and
     whatever the associated ground state $|\s0\>$ is, for which $\H\s |\s0\>=0$ holds, the ground state
     expectation functional has the form
            $$E(f)=\<0|\s e^{\t i\tint \hph(x)\s f(x)\,d^s\!x}\s|\s0\>=e^{\t-\tint\s L[f(x)]\,d^s\!x}\;, $$
            for some suitable choice of the function $L[u]$.

            A canonical representation for the function $L[u]$ is readily determined. We focus on those
            cases that are even functions $L[-u]=L[u]$, which are then real and satisfy $L[0]=0$ and
            otherwise $L[u]\ge0$. Let $f(x)\equiv p\s \chi_\Delta(x)$, where $\chi_\Delta(x)\equiv1$ if $x\in\Delta$ and zero otherwise; moreover, as a modest abuse of notation, we also set $\tint \chi_\Delta(x)\,d^s\!x=\Delta$ as well. Thus
             $$ \<0|\s e^{\t i\tint \hph(x)\s f(x)\,d^s\!x}\s|\s0\>=e^{\t -\Delta\s L[p]}\equiv \int \cos(p\s\l)
             \,d\mu_\Delta(\l) \;, $$
             where we have made use of the symmetry of $L[u]$, and the fact that for each $\Delta>0$ we
             are dealing with a characteristic function (Fourier transform of a probability measure
             $\mu_\Delta$). Thus,
               $$ L[p]=\lim_{\Delta\ra0}\,\Delta^{-1}\,\tint[\s1-\cos(p\s\l)\s]\,d\mu_\Delta(\l)\;.$$
Based on this expression, and assuming convergence, it is clear that the most general function $L[u]$  is
            given by the relation
              $$  L[u]= a\s u^2+\tint_{\l\neq0}[1-\cos(u\s \l)\s]\,d\sigma(\l)\;, $$
              where $a\ge0$ and $\sigma(\l)$ is a nonnegative measure such that
              $$ \tint_{\l\neq0} [\l^2/(1+\l^2)]\,d\sigma(\l)<\infty\;.$$

              The free model solution obtained above is one for which $a=1/(4m_0)$ and $\sigma=0$.
              Let us assume hereafter that $a=0$ and $\sigma\neq0$. Observe that it is possible
              that
                $$ \tint_{\l \neq0}\s d\sigma(\l)=\infty\;, $$
                and in fact this will be the case for the solutions of interest to us
                because we insist that the spectrum of the field operator $\hat{\phi}(x)$ is absolutely
                continuous, and thus for any $\Delta>0$, it is necessary that
                       $$ \lim_{p\ra\infty}\,e^{\t-\Delta\s L[p\s]\s}=0\;.$$

              For   simplicity in what follows, we assume that the measure $\sigma(\l)$ is absolutely
                continuous, and we respect that assumption by setting
                    $$ d\sigma(\l)=c(\l)^2\,d\l\;, $$
                    where $c(\l)$ is known as the ``model function". It has been found that the choice
                    of the model function completely characterizes the ultralocal model under
                    consideration, and, importantly, apart from the free model, all nonlinear
                    ultralocal models are described by the situation where $a=0$ and the model
                    function $c(\l)>0$ \cite{book}.

          \subsubsection*{Model function}
          To ensure that the model function $c(\l)$ has a suitable singularity at $\l=0$, we
          focus our attention on model functions of the form
            $$ c(\l)=(b\s\Delta)^{1/2}\,\frac{e^{\t -y(\l)/2}}{|\l|^\gamma}\;, $$
              where $y(0)=0$, $\gamma=1/2$, and $b$ is a positive constant with dimensions $L^{-s}$. [{\bf Remark:}
               Other $\gamma$ values in the range $1/2<\gamma<3/2$, which are discussed in \cite{book}, can be
               obtained by  suitable, invertible, changes of variables from the case where $\gamma=1/2$.] As a consequence, it follows
              that
                 \b &&E(p)\equiv \<0|\s e^{\t ip\s\s Q}\s|\s0\>\\
                   && \hskip2.4em =e^{\t -(b\s\Delta)\tint[1-\cos(p\s\s\l)]\,\frac{{\t e}^{\t-y(\l)}}{\t|\l|}\,d\l}\\
                   && \hskip2.4em\simeq (b\s\Delta)\int\cos(p\s\s\l)\,\frac{e^{\t-y(\l)}}{|\l|^{1-2b\s\Delta}} \,d\l\;,\e
                   where $Q\equiv\tint \hph(x)\chi_\Delta(x)\,d^s\!x$ and the last relation holds when $0<b\s\Delta\ll1$. Observe that the prefactor $b\s\Delta$ in the last expression is an
                   approximate normalization factor (and an asymptotically correct one!)
                   for the ground state distribution.

                   This latter form of the expectation function for a single degree of freedom
                   readily extends to an infinite set of such fields, with $p=\{p_k\}$ now, such that
                    $$ E_\Delta(p)={\t\prod'}_{k}\,\bigg[ (b\s\Delta)\int\cos(p_k\s\s\phi_k)\,\frac{e^{\t-y(\phi_k)}}{|\phi_k|^{1-2b\s\Delta}}
                     \,d\phi_k\s\bigg]\;.$$
                     Let us consider $\Sigma_k p_k\s\chi_\Delta(x-k\s a)$, where here we have in mind that
                      $\chi_\Delta(x)$ denotes a small hypercubic cell around the origin of area $\Delta=a^s$. As $\Delta=a^s\ra0$ and $\Sigma_k p_k\s\chi_\Delta(x-k\s a)\ra f(x)$, it follows that
                     \b &&\lim_{\Delta\ra0}
                     E_\Delta(p)=E(f)=\<0|\s e^{\t i\tint \hph(x)\s f(x)\,d^s\!x}\s|\s0\>\no\\
                     &&\hskip4.77em =\exp\{-\tint d^s\!x\,\tint[1-\cos(f(x)\s \l)]\,e^{\t-y(\l)}\;d\l/|\l|\}\;.\e

                     This last relation allows us to identify the regularized ground state of a general
                      ultralocal theory as given (with $\hbar$ temporarily restored)  by the expression
                     $$ \Psi(\phi)\equiv{\t\prod}'_k\,(b\s\Delta)^{1/2}
                     \,\frac{e^{\t-y(\phi_k,a,\hbar)/2\hbar}}{\t|\phi_k|^{1/2-b\s\Delta}}
                     \equiv{\t\prod}'_k\Psi_k(\phi_k) \;.$$
                     Given that this expression represents the ground state, it then follows that the regularized Hamiltonian is given by
              \b &&\H_\Delta ={\t\sum'_k}\bigg[ -\half\hbar^2\frac{\t\d^2}{\t\d\phi_k^2}\,a^{-s}+\half\hbar^2
              \frac{\t1}{\t\Psi_k(\phi_k)}\,\frac{\t\d^2\Psi_k(\phi_k)}
              {\t\d\phi_k^2}\,a^{-s}\s\bigg]\\
                 &&\hskip1.8em \equiv -\half\,{\t\sum'}_k \hbar^2\frac{\t\d^2}{\t\d\phi_k^2}\,a^{-s}
                 +{\cal V}(\phi)\;, \e
          where, for the choice of $\Psi(\phi)$ given above,
             \b &&{\cal V}(\phi)\equiv {\t\sum'_k}\,[
             \eigth y'(\phi_k,a,\hbar)^2 -\quarter\s\hbar\s  y''(\phi_k,a,\hbar)+\half\s\hbar\s\gamma_r\, y'(\phi_k,a,\hbar)\,\phi_k^{-1}\\
             &&\hskip6em +\half\s\hbar^2\s\gamma_r(\gamma_r+1)
          \,\phi_k^{-2}]\;; \e
 here
 $$ \gamma_r\equiv \half- b\Delta=\half- b\s a^s\;. $$.

  Consider the pseudofree ultralocal case for which
   $$ y(\phi_k,a,\hbar)=m_0\s \phi_k^2\s a^s\;.$$
   For this choice, it follows that
     $${\cal V}_{pf}(\phi)\equiv \half\s{\t\sum'_k}\,[\s m_0^2\s\phi_k^2\s a^{s}-\hbar\s m_0(1-2\gamma_r)
     +\hbar^2\s\gamma_r(\gamma_r+1)\s\phi_k^{-2}\s a^{-s}\s]\;. $$

Given the Hamiltonian for this case we can immediately determine the lattice action for this
pseudofree ultralocal model. In particular, it follows that
   $$I_{pf}= {\t\sum_k}\{\half[(\phi_{k^\#}-\phi_k)^2\s a^{n-2}+m_0^2\s\phi_k^2\s a^n+\hbar^2\s(\half-b\s a^s)
   (\threebytwo-b\s a^s)\s a^{-2s}\s\phi_k^{-2}\s a^{n}\s]\s\}\;. $$
   In this expression the factor $k^\#$ signifies the next lattice point advanced by one unit
   in the time direction, i.e., if $k=(k_0,k_1,\ldots,k_s)$ then $k^\#=(k_0+1,k_1,\dots,k_s)$.
   Note well that any constant term (zero point energy) in the Hamiltonian cancels out with
   a similar term in the normalization factor in the functional integral and need not be included in the lattice
   action. Observe that the classical limit for which $\hbar\ra0$ accompanied by the continuum
   limit leads to the classical (Euclidean) action for the free ultralocal model.
   \subsubsection*{Interacting ultralocal models}
   Drawing on the foregoing analysis of the pseudofree ultralocal model, we may give a brief discussion
   of interacting ultralocal models. The quartic interaction in the lattice action leads
   to a lattice Hamiltonian of the form
      $$ \H=-\half\hbar^2{\t\sum'_k}\frac{\t\d^2}{\t\d\phi_k^2}+{\cal V}(\phi)\;,$$
      where
      $${\cal V}(\phi)={\t\sum'_k}\,[\s\half m_0^2\phi^2_k\s a^s+\l_0\phi^4_k\s a^s+\half\hbar^2\gamma_r(1+\gamma_r)\phi_k^{-2}\s a^{-s}\s]-E\;. $$
      The constant $E$ is chosen so that the ground state $\Psi(\phi)$ fulfills $\H\s\Psi(\phi)=0$.
      Unfortunately, the form of the expression $y(\phi,a,\hbar)$ that is part of the ground
      state function is unknown, but it surely has the property that as $\l_0\ra0$, then
      $y(\phi,a,\hbar)\ra m_0\s\phi^2\s a^s$ appropriate to the pseudofree model. Stated
      otherwise, the quartic interacting theory is continuously connected to the pseudofree model
      as advertised.

     Although we can not analytically describe the ground state for the quartic ultralocal model,
     we can, as another example, choose  a nonquadratic form for $y(\phi,a,\hbar)$ and see to what interacting model
     it belongs. For example, let us consider
        $$y(\phi,a,\hbar)=m_0\phi^2\s a^s+g_0\phi^4\s a^s\;, $$
     which leads to the potential
        \b &&{\cal V}(\phi)= {\t\sum'}_k\half\{ m_0^2\p^2_k+4m_0g_0\p^4_k +4g_0^2\p^6_k-\half\hbar[m_0(1-2\gamma_r)+2g_0(2\gamma_r-3)\phi^2_k]\\
        &&\hskip5em +\hbar^2
        \gamma_r(1+\gamma_r)\phi_k^{-2}\s\}\s a^s\;. \e

        Evidently this choice describes a model with a mixed quadratic, quartic, and sixth order
        potential. The first three terms -- those without $\hbar$ as a coefficient -- survive
         in the classical limit as $\hbar\ra0$. Again, as the nonlinear coupling $g_0\ra0$, it follows that this interacting model
        is continuously connected to the pseudofree model.
\subsection*{Another Route to Quantize Ultralocal Models}
     Let us now derive the pseudofree ultralocal model by an alternative argument.
     First, we recognize the free model and its ground state on a regularizing lattice as given by
        $$ \Psi_0(\p)=\sqrt{K}\,e^{\t-\half\s m_0\Sigma'_k\p_k^2\s a^s}\;, $$
        which gives rise to the ground state expectation functional
          \b&&E_0(f)=\lim_{\Delta\ra0}\,K\s\int e^{\t i\Sigma'_k p_k\p_k\s a^s-m_0\Sigma'_k\p_k^2\s a^s}
          \,\Pi'_k d\p_k\\
          &&\hskip2.77em =e^{\t-(1/4\s m_0)\tint f(x)^2\,d^s\!x}\;.  \e
          Perturbations in the mass for example would involve expressions of the form
          $$ I_p(m_0)\equiv K\int [\Sigma'_k\p_k^2\s a^s]^p\,e^{\t-m_0\Sigma'_k\p_k^2\s a^s}\,\Pi'_kd\p_k\;,$$
          for which the result is clearly divergent in the continuum limit where the number $N'$
          of spatial lattice points diverges. It is instructive to see just where that $N'$ factor
          originates, and to do so we pass to {\it hyper-spherical coordinates} defined by the
          expressions
           \b &&\p_k\equiv \kappa\eta_k\;,\hskip1em \kappa\ge0\;,\hskip1em -1\le\eta_k\le1\;,\\
                &&\hskip2em \kappa^2\equiv\Sigma'_k\p_k^2\;,\hskip1em 1=\Sigma'_k\eta_k^2\;. \e
          In terms of these variables, it follows that
           $$ I_p(m_0)=2\s K\int [\k^2\s a^s]^p\,e^{\t-m_0\kappa^2\s a^s}\,\kappa^{N'-1}\s d\kappa
           \,\delta(1-\Sigma'_k\eta_k^2)\,\Pi'_kd\eta_k\;. $$
           For large $N'$, this integral may be estimated by steepest descent methods as
            $$ I_p(m_0)=O((N'/m_0)^p)\,I_0(m_0)\;. $$
            Moreover, in a perturbation calculation of $I_1(m_0)$ about $I_1(1)$ (say) it follows that
              $$ I_1(m_0)= I_1(1)-{\delta m_0}\s I_2(1)+\half\s {\delta m_0}^2 I_3(1)-\cdots\;, $$
              where ${\delta m_0}\equiv m_0-1$. Clearly this series is divergent as $N'\ra\infty$, i.e., in the
              continuum limit. Note well that $N'$ makes an explicit appearance in this series
              {\it only} in the factor $\kappa^{N'-1}$ that arises from the measure $\Pi'_k d\p_k$
              put into hyper-spherical coordinates.

              To eliminate those divergences we need to eliminate that appearance of the factor $N'$.
              The only way to eliminate that factor is to change the ground state from that of the
              free system to that of the pseudofree system that takes account of the hard core. To
              attack the hard core directly is difficult and has so far not been a productive direction to
              follow. But, {\it and here is the main point of this discussion}: {\bf To eliminate the
              factor $N'$ that arises from the field measure it suffices to ensure that the
              ground state distribution for the pseudofree theory is such that
                $$ \Psi_{pf}^2(\p)\propto \kappa^{-(N'-R)}\,e^{\t -m_0\Sigma'_k\p_k^2\s a^s} $$
                for some finite parameter $R$}.

                For the ultralocal model, we shall more
                explicitly choose a ground state for the pseudofree model of the form
                $$ \Psi_{pf}(\p)=K' \Pi'_k |\p_k|^{-(1-R/N')/2}\,e^{\t-\half m_0\p_k^2\s a^s}\;,$$
                which leads to the desired form and respects the ultralocal symmetry of the model.
                How do we choose $R$? We require that this expression have an acceptable continuum
                limit, which we study by examining the characteristic function for the ground state
                distribution, i.e.,
                   \b &&\hskip-1.2em E_{pf}(f)=\lim_{a\ra0} {\Pi}'_kK\tint e^{\t ip_k\p_k\s a^s -m_0\p_k^2\s a^s
                   }\,|\p_k|^{-(1-R/N')}\,{\Pi'}_kd\p_k\\
                   &&\hskip2em =\lim_{a\ra0} \Pi'_k\{1-K\tint[1- e^{\t ip_k\p_k\s a^s}]\,
                   e^{\t -m_0\p_k^2\s a^s }\,|\p_k|^{-(1-R/N')}\,\Pi'_kd\p_k\}\;. \e
                   The only way to achieve a meaningful continuum limit is, first, (effectively)
                    choose $m_0=(b\s a^s)\, m$, where $b$ is an arbitrary positive parameter with dimensions of $L^{-s}$, which,
               after a change of variables ($\p_k\ra a^{-s}\l_k$), yields to leading order,
                 \b E_{pf}(f)=\lim_{a\ra0} {\Pi}'_k\{1-K\tint[1- e^{\t ip_k\l_k\s}]\,
                   e^{\t -b\s m\l_k^2\s}\,|\l_k|^{-(1-R/N')}\,\Pi_kd\l_k\}\;, \e
                   and, second, choose $K=c\s (b\s a^s)$ [which fixes $R$ to be $R=2c(b\s a^s)\,N'$],
                    and thus
                   $$E_{pf}(f)=e^{\t-c\s b\tint d^s\!x\tint\{1-\cos[f(x)\s\l]\}\,e^{\t -b\s m\l^2}\,d\l/|\l|}\;.$$
                   Normally, the dimensionless factor $c$ has been chosen as $c=1$ or $c=\half$, but any
                    positive value is acceptable.

                   It is of fundamental importance to observe that we have derived a correct version
                   of the pseudofree ultralocal model by the simple act of choosing the
                   pseudofree ground state distribution to cancel the unwanted factor $N'$, the very factor  that {\bf causes} the divergences in the first place, and then to ensure
                   as meaningful a continuum limit as possible. This simple act ensures that
                   all the moments of interest are now finite and no infinities arise whatsoever.
                   Since this action has the effect of cancelling all divergences, it acts in all
                    necessary ways as would the presumed hard core. In particular, the so-defined,
                     divergence-free interacting theory does not pass continuously to the free
                     theory but instead it passes to an alternative theory, namely, the
                     pseudofree theory. That kind of limiting behavior is the biggest clue to the fact that the
                     interaction acts as a (partial) hard core. Does the simple act of removing the
                      offending factor $N'$ accurately correspond to including the effects of the hard core? In fact, it really doesn't matter if the elimination of the factor $N'$
                    is an accurate realization of the hard core; the putative ``hard core''
                   has already rendered an important service by refocussing our attention beyond those
                   counter terms that are suggested by perturbation theory. Additionally, the study of the soluble ultralocal models has helped us clarify the question of whether removing the
                   factor $N'$ corresponds to accounting for the hard core. Specifically,
                   the solution obtained from a rigorous viewpoint
                   is identical to the one obtained by the supremely simple prescription of choosing a
                   suitable pseudofree model that eliminates the offending factor $N'$. In this sense,
                   the removal of the cause of the divergences, i.e., the factor $N'$, has rendered the theory
                   finite in all respects, and since the result completely agrees with the rigorously obtained
                   result, we are certainly entitled to assert that the removal of the factor $N'$ has
                   accounted for the presence of the hard core in the case of ultralocal models.

                   We shall see that this breathtakingly
                   elementary procedure, coupled with a judicious choice of further details of the
                   pseudofree model, will provide a divergence-free formulation of additional examples
                   of nonrenormalizable models, formulations that would be difficult to arrive at by
                   any other means. It is reasonable that the procedure to eliminate the source of
                   divergences caused by the measure should apply to other models which, in some sense,
                   are ``close'' to ultralocal models. It is also reasonable to expect that traditional
                   nonrenormalizable models are good candidates on which to try a similar approach to
                   deal with otherwise uncontrollable divergences.

\section*{Relativistic Models}
The classical (Euclidean) action for covariant, quartic self interacting scalar fields is given by
   $$ I=\tint\{ \half[({\nabla\p}(x))^2+m_0^2\s\p(x)^2]+\l_0\s\p(x)^4\}\,d^n\!x\;, $$
   for an $n$-dimensional spacetime. To discuss the classical side of the pseudofree situation,
   we recall a classical Sobolev-type inequality (see, e.g., \cite{book}) given by
      $$\{\s\tint \p(x)^4\,d^n\!x\s\}^{1/2}\le c\s\tint[({\nabla\p}(x))^2+m_0^2\s\p(x)^2]\,d^n\!x\;, $$
      which for $n\le 4$ holds with $c=4/3$ and for $n\ge5$, requires that $c=\infty$. This result
      implies that for $n\ge5$, there are fields $\p(x)$ for which the free part of the classical action
      is finite but for which the quartic interaction diverges. These are just the conditions under
      which a classical pseudofree theory different from the classical free theory exists. Thus it is possible
      when $n\ge5$ that the quantum theory also has a pseudofree theory different from its free
      theory.

      We recall that a lattice regularized form of the Euclidean functional integral with only
      two free parameters ($m_0$ and $\l_0$)  has been shown to pass to a (generalized) free
      theory in the continuum limit \cite{aiz}; thus a richer variety of renormalization counterterms is
      required to avoid triviality. Since, for $n\ge5$ the quantum theories are perturbatively
      nonrenormalizable leading to a perturbation series composed of infinitely many
      distinct counterterms, such an approach does not resolve the problem. Our goal is to show
      that an unconventional counterterm suggested by what is needed to remove the source of the
      divergences can lead to a satisfactory resolution of all problems with the relativistic models.
      To that end we now turn our attention to a very different sort of lattice regularized functional
      integral formulation for self interacting relativistic scalar fields.

In particular, relativistic interacting scalar models admit an analogous treatment to that of the ultralocal models,
and in our present discussion we follow reference \cite{prl}.
In begin with, let us  introduce a lattice action defined by the expression
   \b  && I(\p,a,\hbar)\equiv\half{\t\sum_k}\,(\phi_{k^*}-\phi_k)^2\,
 a^{n-2}+\half m_0^2{\t\sum_k} \phi_k^2\, a^n\no\\&&\hskip6em
  + \l_0{\t\sum_k}\phi^4_k\,a^n + \half\s\hbar^2\s{\t\sum}_k {\F}_k(\p)\s a^n \,,\label{t4}\e
where there is an implicit summation over all $n$ nearest neighbors in the positive sense symbolized by
the notation $k^*$, and where the nonclassical counterterm is
 \b &&\hskip-.2cm\F_k(\p)\equiv\frac{1}{4}\s\bigg(\frac{N'-1}{N'}\bigg)^2\s
a^{-2s}\s{\t\sum'_{\s r,\s t}}\s\frac{J_{r,\s k}\s
  J_{t,\s k}\s \p_k^2}{[\S_l\s J_{r,\s l}\s\p^2_l]\s[\S_m\s
  J_{t,\s m}\s\p_m^2]} \no\\
  &&\hskip2cm-\frac{1}{2}\s\bigg(\frac{N'-1}{N'}\bigg)
  \s a^{-2s}\s{\t\sum'_{\s t}}\s\frac{J_{t,\s k}}{[\S_m\s
  J_{t,\s m}\s\p^2_m]} \no\\
  &&\hskip2cm+\bigg(\frac{N'-1}{N'}\bigg)
  \s a^{-2s}\s{\t\sum'_{\s t}}\s\frac{J_{t,\s k}^2\s\p_k^2}{[\S_m\s
  J_{t,\s m}\s\p^2_m]^2}\;. \label{w19}\e
Here,
  \b J_{k,\s
l}\equiv\frac{1}{2s+1}\s\delta_{\s k,\s l\in\{k\s\cup \s
k_{nn}\}}\;, \e where $\delta_{k,l}$ is a Kronecker delta. This latter
notation means that an equal weight of $1/(2s+1)$ is given to the
$2s+1$ points in the set composed of $k$ and its $2s$ nearest
neighbors in the spatial sense only; $J_{k,\s l}=0$ for all other
points in that spatial slice. {\bf [}Specifically, we define
$J_{k,\s l}=1/(2s+1)$ for the points $l=k=(k_0,k_1,k_2,\ldots,k_s)$,
$l=(k_0,k_1\pm1,k_2,\ldots,k_s)$,
$l=(k_0,k_1,k_2\pm1,\s\ldots,k_s)$,\ldots,
$l=(k_0,k_1,k_2,\ldots,k_s\pm1)$.{\bf ]} This definition implies
that $\Sigma'_l\s J_{k,\s l}=1$.

For the ultralocal model, the analog of the constants $J_{k,\s l}$ is
the Kronecker delta, i.e., $\delta_{k,\s l}$. In that case it was important
to respect the physics of the ultralocal model with no interaction between fields at
distinct (lattice) points. For the relativistic models, on the other hand, there
is indeed communication between spatially neighboring points and we can use that
fact to provide a lattice-symmetric, regularized form of the denominator factor. Moreover,
the lack of integrability at $\p_k=0$, for each $k$, which was critical for the ultralocal
models to ensure that the ground state becomes a generalized Poisson distribution in the
continuum limit, is
exactly what is {\it not} wanted in the case of the relativistic models. This latter
fact is ensured by the factors $J_{k,\s l}$ as chosen.

We first focus on our choice of the pseudofree model in the relativistic case, which is chosen
somewhat differently than in the ultralocal case. Specifically, we define the
generating function for the lattice regularized, covariant pseudofree model by
      \b &&\hskip-.3cmS_{pf}(h)=M_{pf}\int \exp[\s Z^{-1/2}\s\Sigma_k h_k\s\p_k\s a^n/\hbar
      -\half{\t\sum_k}\,(\phi_{k^*}-\phi_k)^2\s a^{n-2}/\hbar\no\\
      &&\hskip9em
      -\half\s\hbar\s{\t\sum_k}{\F}_k(\p)\s a^n]
      \,\Pi_k\s d\p_k\,; \e
      here, $Z$ denotes the so-called field strength renormalization constant to
      be discussed below.
Associated with this choice of the pseudofree generating function is
the lattice Hamiltonian for the pseudofree model, which (with the zero point energy subtracted) reads
\b  \H_{pf}= -\half\s{\hbar^2}\, a^{-s}\s{\t\sum'_k}\frac{\t\d^2}{\t\d
\phi_k^2}
  +\half{\t\sum'_k}\,(\p_{k^*}-\p_k)^2a^{s-2}
  +\half\s\hbar^2{\t\sum'_k}\F_k(\p)\,a^s -E_0 \;. \e
  Lastly, we introduce the expression for the pseudofree ground state
    \b  \Psi_{pf}(\p)=
     \sqrt{K} \;\,\s\frac{e^{\t-\Sigma'_{k,l}\p_k\s A_{k-l}\s\p_l\s
     a^{2s}/2\hbar
     -W(\p\s\s a^{(s-1)/2}/\hbar^{1/2})/2}}
     {\Pi'_k[\Sigma'_lJ_{k,l}\s\p_l^2]^{(N'-1)/4N'}}\;,\no\e
     which, in effect, was chosen {\it first}, and then the
     lattice Hamiltonian and the lattice action were derived from it. We discuss the (unknown)
     function $W$ below; however, we observe here that the other factors in $\Psi_{pf}(\p)$
     properly account for both the large field and small field behavior of the ground state.

     In the next section we discuss the continuum limit, and in doing so we are again guided by
     the discussion in \cite{prl}.
\subsection*{Continuum Limit}
     Before focusing on the limit $a\ra0$ and $L\ra\infty$, we
      note several important facts about ground-state averages of the direction
      field variables $\{\eta_k\}$. First, we assume that such averages
      have two important symmetries: (i) averages of an odd number
      of $\eta_k$ variables vanish, i.e.,
      \b \<\eta_{k_1}\cdots\eta_{k_{2p+1}}\>=0\;, \e
      and (ii) such averages are invariant under any spacetime translation, i.e.,
    \b
    \<\eta_{k_1}\cdots\eta_{k_{2p}}\>=\<\eta_{k_1+l}\cdots\eta_{k_{2p}+l}\>\;\e
    for any $l\in{\mathbb Z}^n$ due to a similar translational
    invariance of the lattice Hamiltonian. Second, we note that
    for any ground-state distribution, it is
    necessary that $\<\s\eta_k^2\s\>=1/N'$
for the simple reason that $\Sigma'_k\s\eta_k^2=1$. Hence,
       $|\<\eta_k\s\eta_l\>|\le1/N'$ as follows from the Schwarz
       inequality. Since $\<\s[\s\Sigma'_k\s\eta_k^2\s]^2\>=1$, it
       follows that $\<\s\eta_k^2\s\eta_l^2\s\>=O(1/N'^{2})$.
       Similar arguments show that for any ground-state distribution
         \b  \<\eta_{k_1}\cdots\eta_{k_{2p}}\>=O(1/N'^{p})\;, \e
         which will be useful almost immeadiately.

\subsubsection*{Field strength renormalization}
         For $\{h_k\}$  a suitable spatial test sequence, we insist
         that expressions such as
         \b \int Z^{-p}\,[\Sigma'_k h_k\s\p_k\,a^s]^{2p}\,\Psi_{pf}(\p)^2\,\Pi'_k\s
         d\p_k \label{w20}\e
         are finite in the continuum limit. Due to the intermediate
         field relevance of the factor $W$ in the pseudofree ground state,
         an approximate evaluation of the integral  will
         be adequate for our purposes.
          Thus, we are led to consider
         \b &&\hskip-.3cm K\int Z^{-p}\,[\Sigma'_k
         h_k\s\p_k\,a^s]^{2p}\,\frac{e^{\t-\Sigma'_{k,l}\s\p_k\s A_{k-l}\s\p_l\,
         a^{2s}/\hbar-W}}{\Pi'_k[\s\S_lJ_{k,l}\p_l^2\s]^{(N'-1)/2N'}}\,\Pi'_k\s
         d\p_k\no\\
         &&\hskip.2cm\simeq 2\s K_0\int Z^{-p}\s\k^{2p}\,[\Sigma'_k h_k\s\eta_k\,a^s]^{2p}\\&&\hskip.2cm\times
         \frac{e^{\t-\k^2\s\Sigma'_{k,l}\s\eta_k\s A_{k-l}\s\eta_l\,a^{2s}/\hbar}}
         {\Pi'_k[\s\S_l J_{k,l}\s\eta^2_l\s]^{(N'-1)/2N'}}\,d\k\,\delta(1-\Sigma'_k\eta_k^2)
         \,\Pi'_k\s d\eta_k\;, \label{f5} \no\e
         where $K_0$ is the normalization factor when $W$ is dropped.
         Our  goal is to use this integral to determine a value for
         the field strength renormalization constant $Z$.
         To estimate this integral we first replace two
         factors with $\eta$ variables by their appropriate
         averages. In particular, the quadratic expression in the exponent
         is estimated by
          \b \k^2\s\Sigma'_{k,l}\s\eta_k\s
          A_{k-l}\s\eta_l\,a^{2s}\simeq\k^2\s\Sigma'_{k,l}\s N'^{\,-1}
          A_{k-l}\,a^{2s}\propto \k^2\s N'\s a^{2s}\s a^{-(s+1)}\;, \e
          and the expression in the integrand is estimated by
          \b [\Sigma'_k h_k\s\eta_k\,a^s]^{2p}\simeq\s
          N'^{\,-p}\,[\Sigma'_k h_k\,a^{s}]^{2p}\;. \e
         The integral over $\k$ is then estimated by first rescaling the variable
         $\k^2\ra\k^2/(N'\s a^{s-1}/\hbar)$, which then leads to an overall integral estimate proportional
          to
         \b  Z^{-p}\,[N'\s a^{s-1}]^{\,-p}\,N'^{-p}\,[\Sigma'_k h_k\,a^{s}]^{2p}\;;\e
         at this point, all factors of $a$ are now outside the
         integral.
         For this result to be meaningful in the continuum limit,
         we are led to choose $Z=N'^{\,-2}\s a^{-(s-1)}$. However, $Z$ must
         be dimensionless, so we introduce a fixed positive quantity
         $q$ with dimensions of an inverse length, which allows us to
         set
          \b Z=N'^{\,-2}\s (q\s a)^{-(s-1)}\;. \e
 \subsubsection*{Mass and coupling constant renormalization}
 A power series expansion of the mass and coupling constant terms
 lead to the expressions
  $ \<\s [\s  m_0^2\,\Sigma_k \p_k^2 a^n\s]^p\s\> $ and
  $ \<\s [\s  \l_0\,\Sigma_k \p_k^4 a^n\s]^p\s\> $ for $p\ge1$, which we treat
  together as part of the larger family governed by
  $\<\s [\s  g_{0,r}\,\Sigma_k \p_k^{2r} a^n\s]^p\s\>$ for integral $r\ge1$.
  Thus we consider
  \b &&\hskip-.2cm K\int [\s g_{0,\s r}\Sigma'_k
         \p_k^{2\s r}\,a^s]^{p}\,\frac{e^{\t-\Sigma'_{k,l}\s\p_k\s A_{k-l}\s\p_l\,
         a^{2s}/\hbar-W}}{\Pi'_k[\s\S_lJ_{k,l}\p_l^2\s]^{(N'-1)/2N'}}\,\Pi'_k\s
         d\p_k\no\\
         &&\hskip.7cm\simeq 2\s K_0\int g_{0,\s r}^p\s\k^{2 r p}\,[\Sigma'_k\s\eta_k^{2 r}\,a^s]^{p}\label{ff5}\\
         &&\times\frac{e^{\t-\k^2\s\Sigma'_{k,l}\s\eta_k\s A_{k-l}\s\eta_l\,a^{2s}/\hbar}}
         {\Pi'_k[\s\S_l J_{k,l}\s\eta^2_l\s]^{(N'-1)/2N'}}\,d\k\,
         \delta(1-\Sigma'_k\eta_k^2)
         \,\Pi'_k\s d\eta_k\;.  \no\e
         The quadratic exponent is again estimated as
         \b \k^2\s\Sigma'_{k,l}\s\eta_k\s
          A_{k-l}\s\eta_l\,a^{2s}\propto \k^2\s N'\s a^{2s}\s a^{-(s+1)}\;, \e
while the integrand factor
         \b [\Sigma'_k\eta_k^{2r}]^p\simeq N'^p\s N'^{-rp}\;. \e
        The same transformation of variables used above precedes the integral
        over $\k$, and the result is an integral, no longer depending on $a$, that is proportional to
          \b g_{0,\s r}^p \s N'^{-(r-1)p}\s a^{sp}/N'^{rp}\s a^{(s-1)rp}\;.\e
          To have an acceptable continuum limit, it suffices that
              \b g_{0,\s r}=N'^{(2r-1)}\,(q\s a)^{(s-1)r-s}\,g_r \;,\e
              where $g_r$ may be called the physical coupling factor.
              Moreover, it is noteworthy that
              $Z^r\,g_{0,\s r}=[N'\s (q\s a)^s]^{-1}\,g_r $,
          for all values of $r$, which for a finite spatial volume $V'=N'\s a^s$ leads to a
    finite nonzero result for $Z^r\s g_{0,\s r}$. It should not be a surprise that there are no
          divergences for all such interactions because the source of all divergences has been
          neutralized!

    We may specialize the general result established above to the two cases of
    interest to us. Namely, when $r=1$ this last
    relation implies that $m_0^2=N'\s(q\s a)^{-1}\,m^2$, while when
    $r=2$, it follows that $\l_0=N'^{\,3}\s(q\s a)^{s-2}\s\l$. In
    these cases it also follows that $Z\s m_0^2=[\s N'\s (q\s a)^s\s]^{-1}\s m^2$ and
  $Z^2\s \l_0=[\s N'\s (q\s a)^s\s]^{-1}\s \l$,
    which for a finite spatial volume $V'=N'\s a^s$ leads to
    finite nonzero results for $Z\s m_0^2$ and $Z^2\s \l_0$, respectively.

  \section*{Conclusion}
          For covariant scalar nonrenormalizable quantum field models, we
          have shown that the choice of a nonconventional counterterm, but one
          that is
          still nonclassical,  leads to a formulation for
          which a perturbation analysis of both the mass term and
          the nonlinear interaction term, expanded about the appropriate pseudofree model,
           are term-by-term finite.

           Coupled with the discussion for the
           ultralocal models, it is evident that the present analysis would suggest
           a related formulation for so-called {\it Diastrophic Quantum Field Theories}
           introduced by the author in \cite{dia}. These models are distinguished by the fact
           that they can be viewed as fully relativistic models modified so that some
           (but not all) of the spatial derivatives are dropped; thus these models
           lie, in a certain sense, between the relativistic and ultralocal models.

           It is also hoped that some of these ideas may have relevance in one or more
           formulations of quantum gravity, such as, for example, in the program of
           {\it Affine Quantum Gravity} introduced by the author; see \cite{affine}.

\end{document}